\documentclass[pre,twocolumn,showpacs,preprintnumbers,amsmath,amssymb]{revtex4}

\usepackage[dvips]{graphicx}
\usepackage{dcolumn}
\usepackage{bm}

\begin{document}
\title[Extension of the H\&T model]{Extension of the Hamaneh - Taylor model using the macroscopic polarization for the description of chiral smectic liquid crystals}
\author{H. DHAOUADI, N. BITRI}
\altaffiliation[Also at ]{Centre de Recherches Paul Pascal, 115, Av.
Albert-Schweitzer, 33600 Pessac, France}
\author{S. ESSID, T. SOLTANI, A. GHARBI}
\email{Hassen.Dhaouadi@ipeit.rnu.tn}

\affiliation{ Laboratoire de Physique de la Mati\`ere Molle
Facult\'{e} des Sciences de Tunis, 2092 El Manar TUNIS, Tunisie}

\author{J. P. MARCEROU }
\email{marcerou@crpp-bordeaux.cnrs.fr}
\affiliation{ Centre de Recherches Paul Pascal, 115, Av.
Albert-Schweitzer, 33600 Pessac, France}

\homepage{http://www.crpp-bordeaux.cnrs.fr}

\date{\today}
\begin{abstract}
Chiral smectic liquid crystals exhibit a series of phases, including ferroelectric, antiferroelectric and ferrielectric commensurate structures
as well as an incommensurate $Sm-C^*_\alpha$ phase. We carried out an extension of the phenomenological model, recently presented by M. B.
Hamaneh and P. L. Taylor, based upon the distorted clock model. The salient feature of this model is that it links the appearance of new phases
to a spontaneous microscopic twist : i.e. an increment $\alpha$ of the azimuthal angle from layer to layer. The balance between this twist and
an orientational order parameter $J$ gives the effective phase. We introduce a second orientational order parameter I which physical meaning
comes from the macroscopic polarization, the effect of an applied electric is also studied. We derive new phase diagrams and correlate them to
our experimental results under field showing the sequence of phases versus temperature and electric field in some compounds.

\end{abstract}

\pacs{61.30.Cz}
\maketitle
\section{Introduction}

Chiral smectics are allowed to become ferroelectric and present a helical precession of the optical axes around the layer normal when a tilt ofthe molecules appears in the layers \cite{1}. In the order of decreasing temperature and increasing tilt angle $\theta$, one can observe a
subset of the following full sequence \cite{2,3,4,5} : the smectic A (Sm-A) without tilt angle ($\theta$ = 0) ; the smectic-$C^*_\alpha$
($Sm-C^*_\alpha$) with a tilt angle $\theta$ and an azimuthal angle $\Phi$ precessing with a short incommensurate period along the layer normal
; the smectic-$C^*$ ($Sm-C^*$) with $\Phi$ precessing with a long period and an helicity sign depending on chirality, it is locally
ferroelectric ($P_S \neq$ 0) ; the smectic-$C^*$ Ferri2 ($Sm-C^*_{Fi2}$) where $\Phi $ is periodic over four layers and has a non regular
increment ($\Delta\Phi  \ne 2\pi/4$) within the unit cell, the whole structure shows a long pitch helix with the same sign as the $Sm-C^*$, it
has no macroscopic polarization ($P_S$ = 0) ; the smectic-$C^*$ Ferri1 ($Sm-C^*_{Fi1}$) where $\Phi$ has a non regular increment ($\Delta\Phi
\ne 2\pi/3$) periodic over three layers, a long pitch helix with the opposite sign as in $SmC^*$, it is truly ferrielectric ($P_S\neq$ 0) ; the
smectic-$C^*_A$ ($Sm-C^*_A$) with $\Phi$ periodic over two layers, a regular increment ($\Delta\Phi =\pi$), a long pitch helix with the
opposite sign to $Sm-C^*$, it referred to as antiferroelectric ($P_S = 0$) or anticlinic. Some of these phases may be missing when varying the
chemical formula (tail length) \cite{6} but the order of appearance is conserved. Two of them present a macroscopic polarization, four of them
a long pitch helical precession with a sign change in the middle of the sequence \cite{7,8,9}. Although most of these phases present a
biaxiality of the unit cell, the global structure is uniaxial because of the helical precession around the layer normal and an optical activity
that can be huge results from the rotation of the biaxial structure \cite{10}. Other nomenclatures are also adopted : $Sm-C^* \to Sm-C^*_\beta$, $Sm-C^*_{Fi1} \to
Sm-C^*_\gamma$ and $Sm-C^*_{Fi2} \to AF$ \cite{2,11}. To characterize the different phases several experimental methods can be used : optical
observations, calorimetric measurements and resonant X-ray scattering \cite{3,4}. Other subphases have been proposed \cite{12,13,beta} but are
linked to assumptions which are not accepted unanimously.

Let us briefly mention some theoretical models which have been proposed to describe the structures and behavior of chiral smectic liquid crystals.
The devil's staircase model, also called Ising model because one direction only is allowed for the azimuth, has been proposed soon after the discovery
of the tilted smectic subphases by Chandani et al. \cite{2,12,14}. It is based on the assumption that the competition between the synclinic sequence
(where one layer and the following one have the same azimuth) and the anticlinic one (opposite azimuths) is at the origin of subphases which present
a periodic succession of such sequences. An infinite number of phases with various ratios of synclinic versus anticlinic sequences are predicted
making the so-called devil's staircase \cite{12}. One can define the $q_T$ index as the fraction S/(S+A) of synclinic ordering versus the total number.
Unfortunately the same index applies to phases with different symmetries so it is simply irrelevant. This model is still supported \cite{13} although
it has been ruled out by the results of X-ray resonant scattering experiments \cite{Mach}.

The nearest-neighbors models are based on the definition of a quantity often called $\xi_{\alpha z}$ which measures the tilt inside a layer by the
coordinates of the c-director \cite{PGDG}:

\begin{equation} \left( \begin{array}{c} \xi _{xz} \\ \xi _{yz} \end{array} \right)  = \left( \begin{array}{c} \sin\theta \cos\Phi \\
\sin\theta \sin\Phi \end{array} \right)
\end{equation}

When $\xi_{\alpha z}$ is considered as the macroscopic order parameter, it can be chosen to describe the Sm-A to Sm-C phase transition
\cite{16,17} as its modulus is nearly proportional to the tilt angle $\theta$. If one considers it at the layer level, defining a different
$\xi_{\alpha z}^j $ for each layer j, it can be used to build up a local free energy taking into account the interactions between layers.
Theories have been proposed which deal with these interactions by means of nearest-neighbors couplings $\xi^j  \xi^{j + 1}$ or
next-nearest-neighbors $\xi^j  \xi^{j + 2}$. The form of the free energy is discrete as one has to sum up the interaction terms over all layers
in the integration domain. One can find theories by Sun et al. \cite{18}, Roy et al. \cite{19,20}, Vaupotic \cite{21}. In all cases a suitable
combination of coupling terms could lead to phase diagrams compatible with existing phases. Lorman has introduced linear combinations of this
parameter $\xi _{\alpha z}$ over up to four layers \cite{22} and his treatment leads to the prediction of the well known tilted subphases
except for the Sm-C$^*_\alpha $ phase.

>From an experimentalist's point of view these models are of little usefulness and it is better to stay in the frame of the distorted-clock model as
it gives the best description of the currently encountered phases.

\subsection{The distorted-clock model}

\begin{figure}[ht!]
 \begin{center}
{\includegraphics[width=13cm]{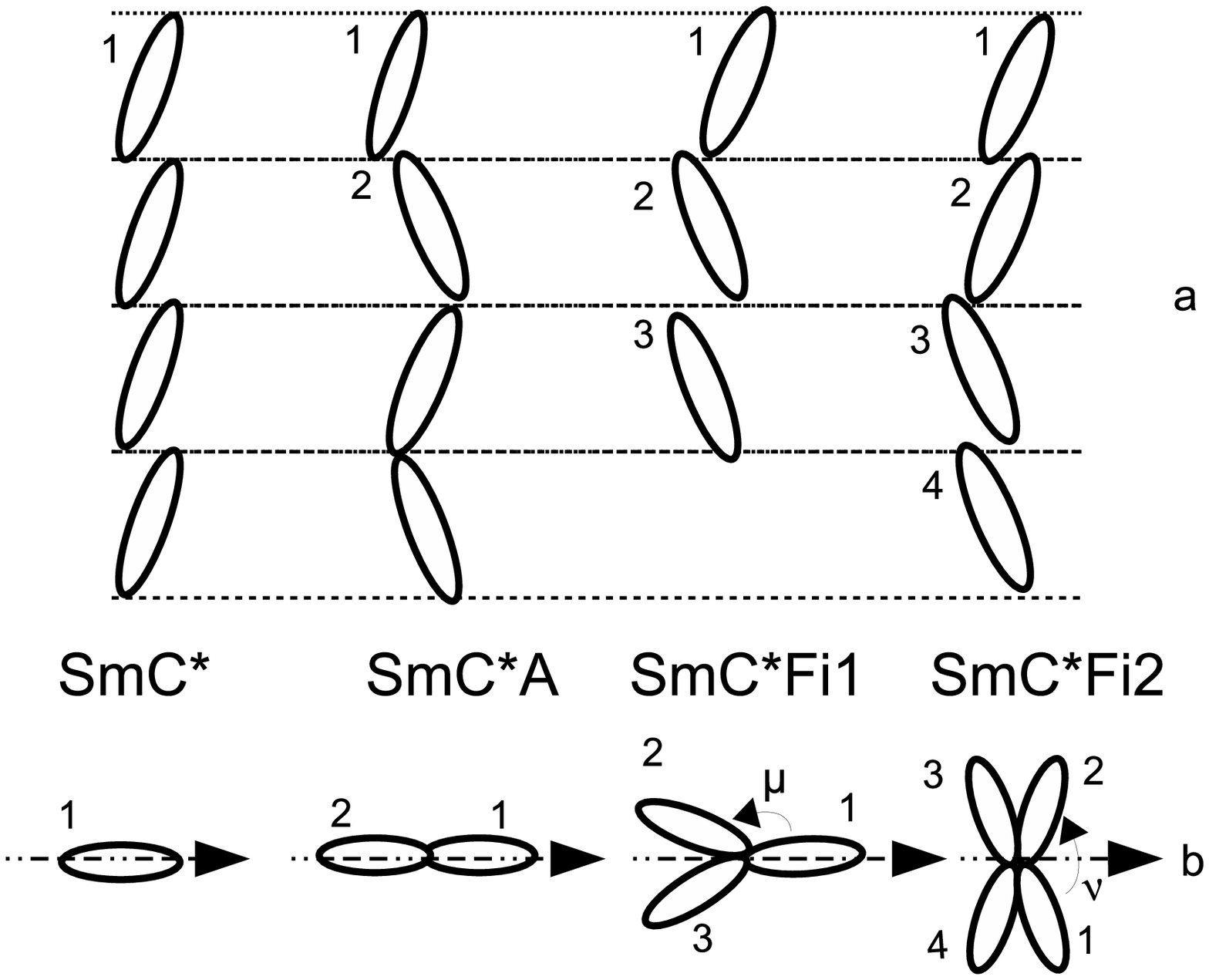}}
 \end{center}
   \caption{Schematic description of the different phases in the distorted clock model. (a) side view (same tilt angle $\theta$)
   period p of 1 to 4 layers, (b) top view (periodic azimuth such that $\Phi_{l+p} = \Phi_{l} + 2\pi$). $\Delta\Phi  = \nu$ or
   $\Delta\Phi  =\pi - \nu$ in the $Sm-C^*_{Fi2}$ and $\Delta\Phi  = \mu$ or $\Delta\Phi = 2(\pi - \mu)$ in the $Sm-C^*_{Fi1}$.
   Arrows indicate the direction $\Phi_0$ taken as the origin of azimuthal angles. Any gap or overlap between molecules is due
   to the hand made drawing and has no physical meaning.}\label{distort}
\end{figure}

This is a purely experimental model describing the observations without ab initio theory, also called XY model because all the azimuthal
directions in the layer plane are allowed. It has been derived by modeling the orientation of the molecules in the layers and then fitting the
resonant scattering experiments with success. With consecutive iterations of the initial regular model \cite{4,Mach}, the authors have
introduced some asymmetries in the azimuthal angle distribution as it is reported in  figure \ref{distort}. The model is still in evolution
concerning the $Sm-C^*_\alpha$  phase \cite{5,CC} but it is coherent and is at the base of the calculations reported here.

\bigskip

\subsection{The Hamaneh-Taylor model}

It is only recently that a new phenomenological way to describe the chiral smectic phases was proposed \cite{HT1,HT2} that we call afterwards the
H\&T model. It is based on the balance between a short range twisting term trying to impose an increment $\alpha$ of the azimuthal angle between
adjacent layers and a long range term linked to the anisotropy of curvature energy in the layer planes. They derived an order parameter
$J=\langle cos 2\Phi_l \rangle$ where the average is taken on the azimuthal angles inside the unit cell. It is non null in the phases
enumerated above and associated to an energy $\eta $ J$^{2}$ where $\eta$ is a coefficient that describes the strength of the long-range
interactions. The short order term reads $\langle cos(\Delta\Phi - \alpha)\rangle$ with $\Delta\Phi = \Phi_l - \Phi_{l-1}$, the elastic term is
$\eta $ J$^{2}$ and the free energy of the sample is~:

\begin{equation}
\label{eq1}
\frac{F}{F_0}  =  \left\langle \cos \left( \Delta\Phi-\alpha \right) \right\rangle + \eta J^2
\end{equation}

The order of magnitude of $F_0$ is the electrostatic energy ($- P_S \cdot E_c$) necessary to drive at the field $E_c$ the phase transition to a
ferroelectric phase with a polarization $P_S$ \cite{HT1,HT2} while $\eta$ is of order unity.

This leads to a phase diagram in the ($\eta, \alpha$) plane showing the sequences of subphases which can be observed in a given liquid crystal.
This model presents some limitations. First, it introduces a phase with six layers which was never observed experimentally. Second, the extent
of the three layers phase is very small.

After this review we introduce in the next sections a new orientational order parameter I that will describe the contribution of the macroscopic
polarization $P_S$ to the ordering. We present then the phases diagrams obtained from a numerical calculation. Eventually we compare the theoretical
results with our experimental data obtained on several compounds

\section{The orientational order parameter}

In H\&T model it is shown clearly that the average $J=\langle cos 2 \Phi_l \rangle$ is non zero in all the phases described by the distorted
clock model, by analogy we state that another average $I=\langle cos \Phi_l \rangle$ is also non null in the phases like $Sm-C^*$ and
$Sm-C^*_{Fi1}$ which possess a macroscopic polarization. The origin $\Phi_0$ is such that the averages over sine functions are zero
\cite{HT1,HT2}, it will be the azimuth $\Phi_1$ of the first layer except in the $Sm-C^*_{Fi2}$ phase where it is equal to $\Phi_1 +
\upsilon/2$. We get the following table where $\mu$ and $\upsilon$ stand for the characteristic angles of asymmetry in the Ferri phases :

\begin{table}[h]
\begin{tabular}
{|c|c|c|c|}\hline
 &\ I = $\langle cos \Phi_l \rangle$&\ J = $\langle cos  2\Phi _l \rangle$& origin $\Phi_0$\\
\hline
$Sm-C^*_\alpha$ & 0 & 0 & X\\
\hline
$Sm-C^*$ & 1 & 1 & $\Phi_1$\\
\hline
$Sm-C^*_A$& 0 & 1 & $\Phi_1$ \\
\hline
$Sm-C^*_{Fi1}$ & $[1 + 2  cos \mu]/3$ & $[1+ 2  cos 2\mu]/3$ & $\Phi_1$\\
\hline
$Sm-C^*_{Fi2}$ & 0 & $-cos \upsilon$ & $\Phi_1 + \upsilon/2$\\
\hline
\end{tabular}
\caption{order parameters I and J and origin of angles in the different  phases}\label{tableau}
\end{table}

\subsection{introduction of an $I^2$ term in the free energy }

The symmetry argument of R. Meyer \cite{1} stating that there exists a polarization $P$ as soon as the the layer normal $\overrightarrow{N}$ and the
director $\overrightarrow{n}$ make an angle $\theta$ can be translated by introducing in the free energy the mixed product of $\overrightarrow{P}$,
$\overrightarrow{N}$ and $\overrightarrow{n}$. Taking into account the table \ref{tableau}, the angle between $\overrightarrow{N}$ and
$\overrightarrow{n}$ reads $I \theta$, so with the addition of the self energy of the polarization, one gets :

\begin{equation}
\label{eq2}
\Delta F_{P}   =  \frac{P^{ 2}} {2\varepsilon_0\chi}  -  C P   I  \theta
\end{equation}

by minimizing over P one finds

\begin{eqnarray}
\label{eq3}
P_S  &=&  \varepsilon_0  \chi  C I \theta \nonumber \\
\\
\Delta \tilde {F}_P   &=&  - \frac{P_S^{ 2}} {2\varepsilon_0\chi} =  - \frac{\varepsilon _0  \chi  C^2 \theta
^2 }{2}I^2 \nonumber
\end{eqnarray}

We have thus demonstrated that the term due to the macroscopic polarization $P_S$, present only when I is non zero, can be written $\tilde{\gamma} \theta^2  I^2$ and we can add it to H\&T free energy after a little bit of algebra on the orientational order parameter.

\subsection{relationship between $I^2$ and $J^2$ terms.}

Let us start from the de Gennes orientational order parameter in the Sm-A phase, in what follows one considers the z axis to be perpendicular
to the smectic layers.

\begin{equation}
 Q^0_{ij}   =  n_i  n_j - \frac{1}{3}\delta _{ij}   =
 \left( {{\begin{array}{c c c}
 -1/3  & 0  & 0  \\
 0  & -1/3  & 0  \\
 0  & 0 & +2/3
\end{array} }} \right)
\end{equation}

One can build up an orientational order parameter $Q_{ij}$ for all the tilted phases of the distorted clock model by first computing for each layer,
in an axis frame where z is the layer normal, the expression of $Q^0_{ij}$ after a tilt of the director $\vec{n}$ by an angle $\theta$.
This rotation is made around an axis inside the layer such that $\Phi$ is the angle between the c-director (projection of the director $\vec{n}$
in the layer plane) and the x axis.
\begin{widetext}
 \begin{eqnarray}
\nonumber  Q^{\,\theta \,\Phi }_{ij} = \left( 1 - \frac{3}{2}\,\sin^2\theta \right) \left( \begin{array}{c c c}
 -1/3  & 0  & 0  \\
 0  & -1/3  & 0  \\
 0  & 0 & +2/3
\end{array} \right) &+& \frac{1}{2}\sin^2\theta
\left( \begin{array}{c c c}
 \cos 2\Phi  & \sin 2\Phi  & \ 0\   \\
 \sin 2\Phi  & -\cos 2\Phi  & \ 0\   \\
 0  & 0 &\  0\
\end{array} \right)\\
\\ \nonumber
 ~ &-& \sin \theta \cos \theta
 \left( \begin{array}{c c c}
  0 & 0  & \ \cos \Phi\   \\
 0  & 0  & \ \sin \Phi\   \\
 \cos \Phi  & \sin \Phi &\  0\
\end{array} \right)
\end{eqnarray}
\end{widetext}

Finally  $Q_{ij}$ is computed by averaging over the unit cell of each phase. For the $Sm-C^*_\alpha$ the average over $\Phi$ is null for the
second and the third matrix and there remains :

\begin{equation}\label{equalpha}
    Q^{\alpha }_{ij} = \left( 1 - \frac{3}{2}\,\sin^2\theta \right)  \left( \begin{array}{c c c}
 -1/3  & 0  & 0  \\
 0  & -1/3  & 0  \\
 0  & 0 & +2/3
\end{array} \right)
\end{equation}

For all other phases, one can write a general formula for $Q_{ij}$ which is a function of $\Phi_0$ defined in table \ref{tableau} as the angle
between the origin of azimuthal angles in the unit cell and the x axis. The resulting order parameter $Q_{ij}$ is unique ; it is only its
expression in a given frame which depends on $\Phi_0$. One readily finds that the order parameters $J=\langle cos 2\Phi_l \rangle$ and
$I=\langle cos \Phi_l \rangle$ can be factorized in the last two matrices. For example in the $Sm-C^*_{Fi1}$ phase where $\Phi_1=\Phi_0,
\Phi_2=\Phi_0 + \mu, \Phi_3=\Phi_0 - \mu$, the averages in the unit cell give $\left(\cos\Phi_1 + \cos\Phi_2 + \cos\Phi_3 \right) = 3 I
\cos\Phi_0$, $\left(\cos2\Phi_1 + \cos2\Phi_2 + \cos2\Phi_3 \right)= 3 J \cos2\Phi_0$, $\left(\sin\Phi_1 + \sin\Phi_2 + \sin\Phi_3 \right) = 3
I \sin\Phi_0$, $\left(\sin2\Phi_1 + \sin2\Phi_2 + \sin2\Phi_3 \right)= 3 J \sin2\Phi_0$ and so on for all the phases :
\begin{widetext}
\begin{eqnarray}
  Q_{ij} = \left( 1 - \frac{3}{2}\,\sin^2\theta \right)  \left( \begin{array}{c c c}
 -1/3  & 0  & 0  \\
 0  & -1/3  & 0  \\
 0  & 0 & +2/3
\end{array} \right) &+& \frac{J}{2} \sin^2\theta
\left( \begin{array}{c c c}
 \cos 2\Phi_0  & \sin 2\Phi_0  & \ 0\   \\
 \sin 2\Phi_0  & -\cos 2\Phi_0  & \ 0\   \\
 0  & 0 &\  0\
\end{array} \right) \nonumber\\
\\ \nonumber
  &-&   I \sin \theta \cos \theta
 \left( \begin{array}{c c c}
  0 & 0  & \ \cos \Phi_0\   \\
 0  & 0  & \ \sin \Phi_0\   \\
 \cos \Phi_0  & \sin \Phi_0 &\  0\
\end{array} \right)
\end{eqnarray}
\end{widetext}
It is straightforward to remark that the only parameters that should be retained for building the free energy are the factorized matrix
coefficients $J\, \sin^2 \theta \sim \theta^2 J$ and $I  \sin\theta\cos\theta \sim \theta I$. So the H\&T term should read $\tilde{\eta}
\theta^4  J^2$ and the polarization term as demonstrated above $\tilde{\gamma}  \theta^2   I^2$. We took advantage of this by considering
that we could write the modified H\&T free energy under the form :
\begin{eqnarray}\label{avecP}
  \frac{F}{F_0}  &=&  \left\langle \cos \left( \Delta\Phi-\alpha \right) \right\rangle + \tilde{\eta}  \theta^4  J^2 + \tilde{\gamma}
  \theta^2  I^2 \\
   \nonumber &=&  \left\langle \cos \left( \Delta\Phi-\alpha \right) \right\rangle + \eta  J^2 + \gamma  \sqrt{\eta}  I^2
\end{eqnarray}

We then assume that the temperature dependence of the $\eta$ coefficient is due to $\theta^4  \sim  (T_c -T)^2$, where $T_c$ is the temperature
of appearance of the tilt angle and that the coefficient $\gamma$ depends only on the compound and not on the temperature. We eventually build
phase diagrams in the ($\eta, \alpha$) plane, each one for a different value of $\gamma$ taking it to be of order unity, from 0 to 1 (see e. g.
figure \ref{gamma}). These results show that the $Sm-C^*$ and $Sm-C^*_{Fi1}$ domains grow with $\gamma$ i.e. with the permanent polarization,
let us consider now what happens in the presence of an external electric which is known to induce phase transitions to polar phases \cite{24}.

\subsection{effect of electric field in the layer plane}

An electric field applied to the sample always creates a small dielectric polarization proportional to it which is the same in all studied phases to
first approximation. But when there is already a spontaneous polarization $P_S$ it will displace the energy by a term which reads roughly
$- P_S\cdot E$. The free energy can be written by slightly modifying equation(\ref{eq2}):

\begin{equation}
\label{eq5}
\Delta F_E   =  \frac{P^2}{2\varepsilon _0 \chi }  -
 C P I \theta   -  P\cdot E  -  \frac{\varepsilon_\bot  E^2}{2}
\end{equation}

by minimizing over P, one finds:
\begin{eqnarray}
\label{eq6}
\tilde{P}  &=&  \varepsilon_0  \chi  \left(   C \theta  I +  E \right)  =  P_S   + \varepsilon_0  \chi  E
\nonumber\\
\Delta \tilde {F}_E  &=& - \frac{P_S^2}{2\varepsilon _0 \chi }  -  P_S\cdot E  - \frac{\widetilde{\varepsilon}_\bot  E^2}{2}
\end{eqnarray}

For a given value of the electric field, the third term $-\widetilde{\varepsilon}_\bot E^2/2$ is the same for all the phases and does not depend on
the orientational order parameters I and J so we simply forget about it. The first two terms have the same order of magnitude $F_0\sim - P_S\cdot E_c$
and are respectively quadratic and linear versus $\theta$, so by taking $\eta$ as the main parameter we can write down :

\begin{equation}
\label{eq7}
\frac{F}{F_0 }  =  \langle \cos \left( \Delta\Phi - \alpha  \right)\rangle   +  \eta J^2  +  \gamma \sqrt \eta  I^2  +   \delta \sqrt[4]{\eta} I
\end{equation}

\begin{figure*}[ht]
\includegraphics[width=16cm]{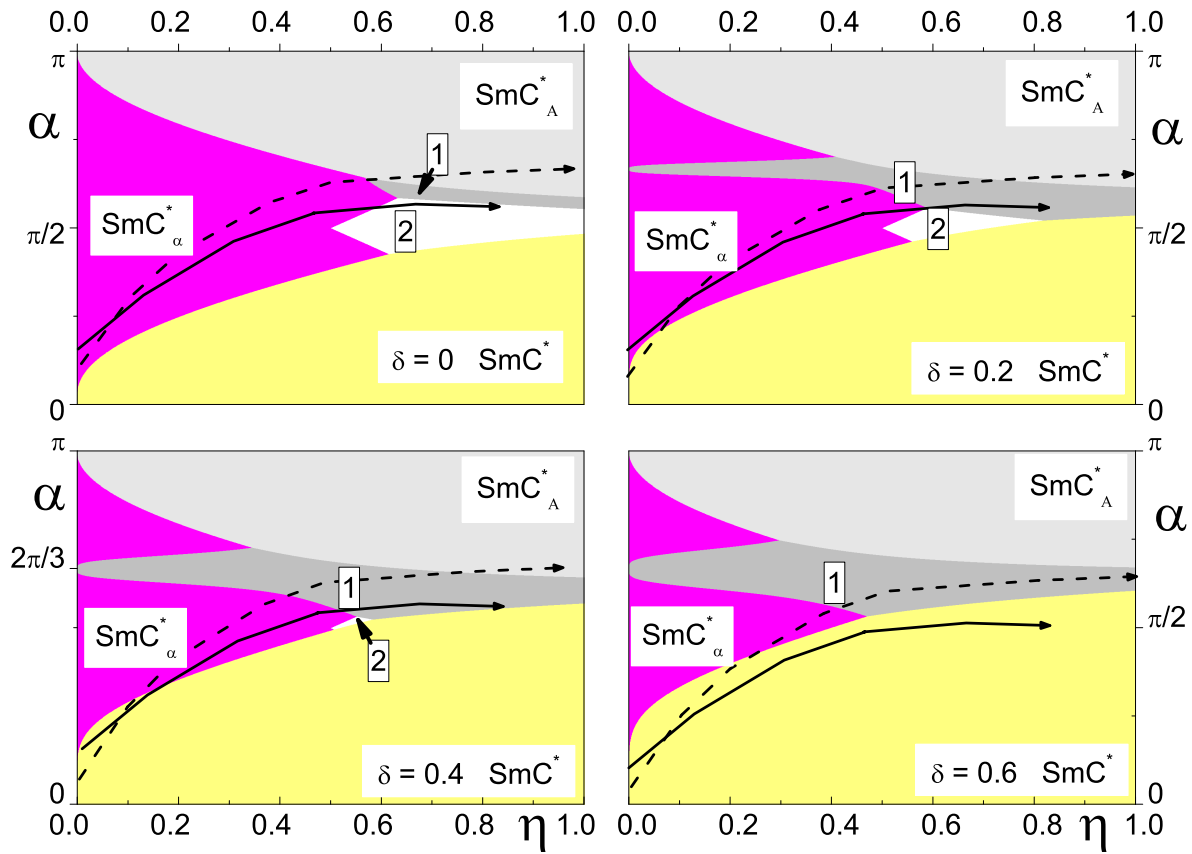}
\caption{(Color online) Diagrams obtained with applied field for $\gamma = 0.2$ and $\delta$ = 0, 0.2, 0.4 and 0.6. The black plain curve corresponds
to an estimate of the path $\eta(\alpha)$ followed by C10F3 and the dotted line to $\eta(\alpha)$ for C7F2. }\label{delta}
\end{figure*}

We use this expression to compute the phase diagrams in the ($\eta, \alpha$) plane, each one for a different value of $\gamma$ and $\delta$, see e.g.
figure \ref{delta}.

\section{Numerical study and phase diagrams}

\begin{figure*}[ht]
\includegraphics[width=16cm]{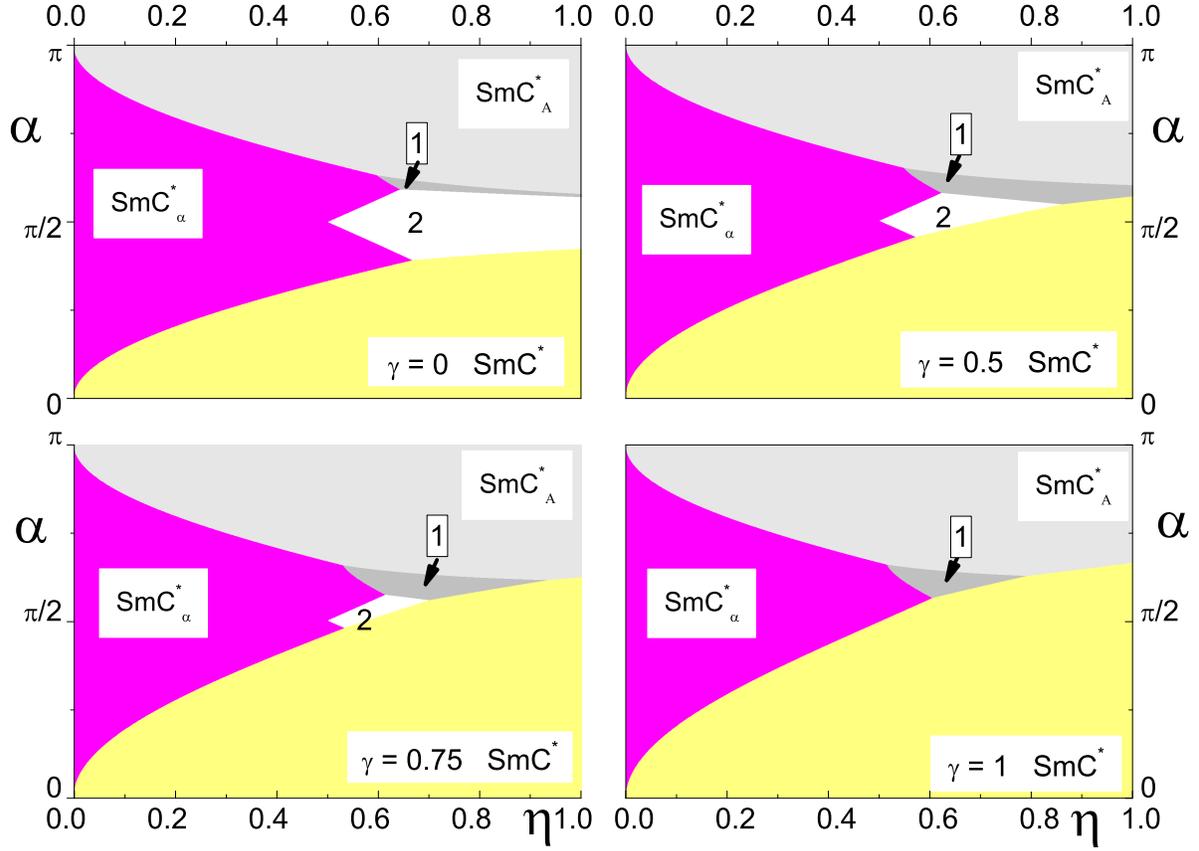}
\caption{(Color online) Ground states diagram for $\gamma$ = 0, 0.5, 0.75 and 1. The symbols 1 and 2 stand for ferri1 and ferri2.}\label{gamma}
\end{figure*}

In the different phases of the distorted clock model represented on the figure \ref{distort}, we establish at first the expression of the free energy
for a given ($\eta $,$\alpha )$ by computing the order parameters I and J as well as the quantity $\langle cos(\Delta\Phi - \alpha) \rangle$ for each
structure. The H\&T model predicts the existence of a phase with six layers, we are going to disregard this phase on the one hand by the fact that it
was never observed on the other hand because it disappears of the diagram once the term due to the polarization $P_S$ is added. Other structures not
observed experimentally have been briefly tested like a four layers asymmetric phase which has a less favorable energy than the Ferri2 phase. Let us
point out that as $F_0$ is negative we look for an absolute maximum of $F/F_0$ to get the best phase.

\begin{itemize}
\item
 In the $Sm-C^*_\alpha$ phase the short order term reduces to 1 while the additional terms vanish. J = 0 and I = 0 thus the free energy is F = F$_0$

\item In the $Sm-C^*$ phase J = 1 and I = 1 thus

\begin{equation}
\label{eq8}
\frac{F}{F_0}  =  \eta   + \cos \alpha   + \gamma \sqrt \eta   + \delta \sqrt[4]{\eta}
\end{equation}

\item In the $Sm-C^*_A$ phase J = 1 and I = 0 thus

\begin{equation}
\label{eq88}
\frac{F}{F_0 }  =   \,\eta   -  \cos\alpha
\end{equation}

\item In the $Sm-C^*_{Fi2}$ phase  $J = - \cos \upsilon$, I = 0 and by minimizing F over $\upsilon$ we find for $\eta > 0.5$ that the preferred angle
is such that $\sin \tilde{\upsilon} = \sin \alpha /2\eta$ and the free energy reads~:

\begin{equation}
\label{eq9}
\frac{F}{F_0}  =  \eta   +  \frac{\sin^2 \alpha }{4 \eta}
\end{equation}

\item In the $Sm-C^*_{Fi1}$ phase $J = \left( 1 + 2\cos 2\mu \right)/3$, $I = \left( 1 + 2\cos \mu \right)/3$  so all the terms of equation(\ref{eq7})
must be explicited :
\begin{eqnarray}
\label{eq10}
  \nonumber \langle \cos(\Delta\Phi - \alpha) \rangle &=& \left( 2 \cos(\mu-\alpha)  + \cos (2\mu + \alpha)\right) / 3\\
  \nonumber \eta  J^2  &=& \eta \left( 4 \cos^2 \mu - 1 \right)^2 / 9\\
  \nonumber \gamma  \sqrt \eta  I^2  &=& \gamma  \sqrt \eta   \left( 2 \cos \mu + 1 \right)^2 / 9\\
  \delta \sqrt[4]{\eta} I  &=& \delta \sqrt[4]{\eta}  \left | 2 \cos \mu + 1 \right | / 3
\end{eqnarray}

The maximization of the energy has to be made numerically giving the preferred value of $\mu$ and $F/F_0$;
\end{itemize}

\subsection{Computation of phase diagrams}
Let us first discuss the physical meaning of these diagrams. On the x-axis one reports the values of the parameter $\eta$ that we take in mean
field approximation as being the fourth power of the tilt angle $\theta$, thus the second power of the distance in temperature from the Sm-A
phase - i.e. the tilt angle appears at the Sm-A to Sm-C phase transition and follows a $\sqrt{T_c - T}$ law. So the x-axis represents
decreasing temperatures from the Sm-A at left to the right. The y-axis shows the scale of variation of the ad-hoc angle $\alpha$ from 0 to
$\pi$. This parameter makes the richness of the H\&T model, it means physically that due to the chirality the director wants to be twisted from
layer to layer by the angle $\alpha$ and it is the balance between this tendency and a more uniform azimuthal angle distribution measured by
the I and J order parameters that gives rise to the effective structures. It is remarkable that with so few parameters one can get all the
structures determined experimentally before the theory was developed by H\&T. One is not as usual trying to fit the experiment to a theory
imposed ab initio.
\begin{figure}
 \begin{center}
{\includegraphics[width=8cm]{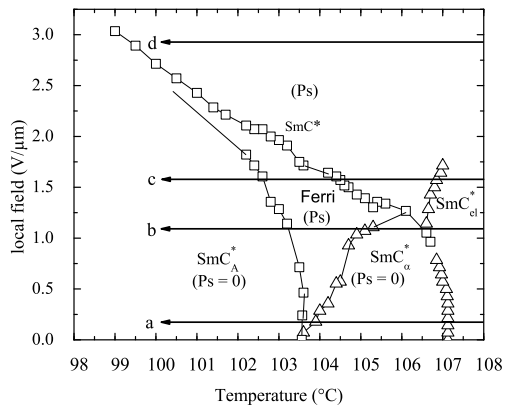}}
 \end{center}
   \caption{(E,T) phase diagram of the compound C7F2. The constant field paths a to d correspond to increasing values of the parameter
   $\delta$.}\label{C7F2}
\end{figure}
What we introduce in this paper are new terms linked to the polar nature of two phases in the tilted chiral smectics nomenclature, the $Sm-C^*$
and the $Sm-C^*_{Fi1}$. We express these terms as functions of the x-axis parameter, one measuring the spontaneous polarization and the other
its coupling with external field. Our aim is to show that the domain filled in by these two phases will be extended in the diagram.

\begin{figure}
 \begin{center}
{\includegraphics[width=8cm]{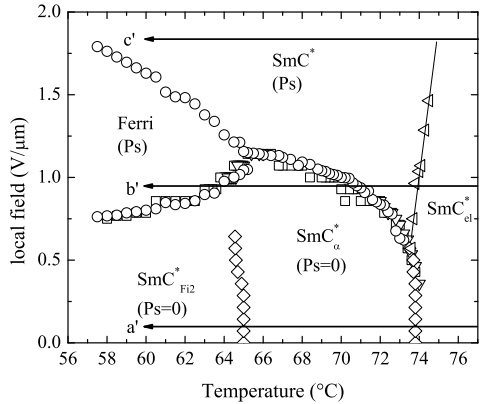}}
 \end{center}
   \caption{(E,T) phase diagram of the compound C10F3. The constant field paths a' to c' correspond to increasing values of the parameter
   $\delta$.}\label{C10F3}
\end{figure}
We first present in figure \ref{gamma} some diagrams obtained for different values of $\gamma$ that correspond to the ground states without
applied field. As $\gamma$ increases one notices an expansion of the domains corresponding to the $Sm-C^*$ and $Sm-C^*_{Fi1}$ phases which are
the only ones with a spontaneous polarization. It is to be noticed that for $\gamma = 1$ the domain corresponding to $Sm-C^*_{Fi2}$ has
completely disappeared.

The diagram obtained with $\gamma $ = 0.2 is chosen to illustrate qualitatively (figure \ref{delta}) the sequences of phases which appear for the two
compounds we have studied experimentally in our group : the C10F3 and the C7F2 \cite{6,24,25,C7F2}.

With the applied field i.e. the parameter $\delta$ one notices also an expansion of the $Sm-C^*$ and $Sm-C^*_{Fi1}$ domains. Furthermore it
appears a band corresponding to $Sm-C^*_{Fi1}$ in full centre of the alpha domain, this band widens gradually as $\delta$ increases.

\subsection{Comparison with experiment}

In the figure \ref{delta} are represented the estimated paths followed on decreasing the temperature by two compounds studied experimentally in
our group : the C7F2 (dotted curve) and C10F3 (plain). The paths are unchanged in the different panels of the figure as they depend only on the
temperature, the nature of the phase encountered at a given point changes as the $Sm-C^*_{Fi1}$ and $Sm-C^*$ domains grow.

The C7F2 compound shows the following ground phase sequence: $SmA
\rightarrow Sm-C^*_\alpha \rightarrow Sm-C^*_A$. The corresponding
path (dotted line in fig. \ref{delta}) has to include the
($Sm-C^*_\alpha) - (Sm-C^*_{Fi1}) - (Sm-C^*_A)$ triple point (below
arrow a in fig. \ref{C7F2}) as a very weak field reveals the
$Sm-C^*_{Fi1}$ phase. The increase in the electric field enlarges
the $Sm-C^*$ and $Sm-C^*_{Fi1}$ domains. At a certain point the path
crosses the $Sm-C^*$ domain then the $Sm-C^*_\alpha$ and the
$Sm-C^*_{Fi1}$  finishing in the $Sm-C^*_A$ (b and c in fig.
\ref{C7F2}). For a large enough field the $SmC^*$ domain will cover
almost all the length of the path (arrow d).

The C10F3 compound presents at zero field the following phase
sequence: $Sm-A \rightarrow Sm-C^*_\alpha  \rightarrow
Sm-C^*_{Fi2}$. For weak electric field (arrow a' in figure
\ref{C10F3}) the sequence remains almost the same. For a higher
electric field (arrow b') the curve corresponding to the C10F3
begins in the $Sm-C^*$ domain then passes in the $Sm-C^*_\alpha$
before the $Sm-C^*_{Fi1}$ grows bigger at the expense of the
$Sm-C^*_{Fi2}$ domain. One then encounters successively two triple
points : the ($Sm-C^*_\alpha) - (Sm-C^*_{Fi1}) - (Sm-C^*_{Fi2})$ and
the ($Sm-C^*) - (Sm-C^*_\alpha) - (Sm-C^*_{Fi1})$. For a strong
electric field the $Sm-C^*$ phase is dominant (arrow c').

\subsection{Discussion}

Another comparison to experiment can be made with the published values of the distortion angles $\mu$ and $\upsilon$.  A. Cady et al. \cite{3}
find for the $Sm-C^*_{Fi2}$ structure a value of $\upsilon$ of about $164^\circ $, Roberts et al. \cite{Gleeson}  measured the angular
distortion of the $Sm-C^*_{Fi2}$ and  $Sm-C^*_{Fi1}$ structures in mixtures at two temperatures, they found a value of $\upsilon$ of about
$166^\circ $ and that of $\mu$ of about 152 to $160^\circ $ with no discernable dependence on temperature. Starting from our equations we made
the calculation of $\upsilon$ and $\mu$ for different points from the two curves of figure \ref{delta}. We found values of $\mu$ varying from
156 to $161^\circ $ for both compounds (under field) which is comparable to that reported by Cady, Roberts et al. \cite{3,Gleeson}. On the
other hand the values we computed for the $\upsilon$ angle, about $142^\circ $, are slightly lower than the measured values \cite{Gleeson,CC2}
but still far from the regular clock model ($\upsilon = 90^\circ $).

We have shown that taking into account the macroscopic polarization in H\&T theory one is led to an expansion of the $Sm-C^*$ and
$Sm-C^*_{Fi1}$ existence domains which is correlated with the (E-T) phase diagrams we have obtained experimentally.

As noticed by H\&T the translation of the theory in a quantitative way requires the knowledge of the physical path $\alpha(\eta)$ or separately
$\alpha(T)$ and $\eta(T)$. What we add is the requirement for the new $\gamma$ and $\delta$ parameters which can be derived from the
measurement of $P_S$ and E (see e.g. equation \ref{eq6}). The parameter $\eta$ should be considered to be zero at the temperature $T_c$ where
the tilt angle appears and to follow a $(T_c - T)^2$ law below. In the H\&T frame it gets remarkable values at the phase boundaries like $\eta
= 1 + \cos \alpha$ at the $Sm-C^*_\alpha$ to $Sm-C^*_A$ phase transition. So the measurement of $\alpha$ in the $Sm-C^*_\alpha$ phase is
required.

Several experimental methods have been reported to measure the parameter $\alpha$, let us quote the resonant X-rays diffraction which allows to
measure a periodicity ranging usually from eight to three layers in $Sm-C^*_\alpha$ with a decrease (i.e. an increase of $\alpha$) when cooling
down the sample \cite{5,Mach,CC,tinh}. However there are a few exceptions where the periodicity varies from about ten to fifty layers ($\alpha$
is decreasing) when lowering the temperature in the assumed $Sm-C^*_\alpha$ range \cite{4,tinh}. This is not obviously an experiment accessible
to everyone. Differential optical reflectivity has been also used to acquire the temperature variation of the helical pitch in the
$Sm-C*_\alpha$ phase and has also shown an increase of $\alpha$ when cooling down \cite{ellip}. Another method used by Isaert \cite{Isaert}
consists in measuring the spacing of the Friedel fringes which appear in the reflected light texture on the free surface of drops, this simple
method could allow a fast measurement of $\alpha$. Finally another candidate is the gyrotropic method given by Ortega and al \cite{26} who by a
measurement of the ellipticity of eigenmodes claim to get the pitch and thus the angle $\alpha$.

\section{ Conclusion}

In this paper we report a new successful method for the description of chiral smectic liquid crystals based on the Hamaneh - Taylor model. The
introduction of the polarization and the electric field contributions give results that sound in good agreement with experiment and allow
explaining the appearance under field of an intermediate ferrielectric phase. The measurement of the parameter $\alpha$ and of the polarization
$P_S$ should allow to trace the paths followed by a given compound in the $(\eta , \alpha )$ plane. In the future we plan to introduce the
helicity and its sign evolution in the phase sequence.

\begin{acknowledgements}
We wish to acknowledge the support of CMCU grant \# 06/G 1311 and CNRS/DGRSRT grant \#~18476.
\end{acknowledgements}


\bibliography{HT3}

\end{document}